\documentclass[acmsmall,screen]{acmart}

\setcopyright{none}
\renewcommand\footnotetextcopyrightpermission[1]{}
\acmConference[FSE '27]{The ACM International Conference on the Foundations of Software Engineering}{July 12--16, 2027}{Shenzhen, China}
\acmYear{2027}
\copyrightyear{2027}
\acmDOI{}
\acmISBN{}

\usepackage{booktabs}
\usepackage{xspace}
\usepackage{enumitem}
\usepackage{amsmath}
\usepackage{multirow}
\usepackage{tikz}
\usetikzlibrary{arrows.meta,positioning}

\pdfoutput=1

\newcommand{\system}{SecTDD\xspace}
\newcommand{\joint}{Hidden Secure-Pass@1\xspace}
\newcommand{\func}{Hidden Func@1\xspace}
\newcommand{\secure}{Hidden Secure@1\xspace}
\newcommand{\tfor}{TFOR\xspace}
\newcommand{\PrimaryTrajectories}{2,705}
\newcommand{\PrimaryTaskInstances}{31}
\newcommand{\PrimaryCWEs}{16}
\newcommand{\UpfrontMacroDelta}{19.3}
\newcommand{\UpfrontPositive}{7}
\newcommand{\UpfrontNegative}{2}
\newcommand{\MRepairs}{80}
\newcommand{\MRegressions}{0}
\newcommand{\RawRepairs}{83}
\newcommand{\RawRegressions}{3}
\newcommand{\StructuredWins}{6}
\newcommand{\StructuredLosses}{6}
\newcommand{\StructuredTies}{453}
\newcommand{\MinimumHolmP}{0.125}

\begin{document}

\title{Security Tests as Executable Specifications for LLM Code Generation:\\Benefits, Trade-offs, and Coverage Limits}

\author{Yunhao Liang}
\affiliation{%
  \institution{Chengdu Institute of Computer Applications, Chinese Academy of Sciences and University of Chinese Academy of Sciences}
  \country{China}
}

\author{Chengguang Gan}
\affiliation{%
  \institution{Independent Researcher}
  \country{Japan}
}

\author{Ruixuan Ying}
\affiliation{%
  \institution{Institute of Multidisciplinary Research for Advanced Materials (IMRAM), Tohoku University}
  \country{Japan}
}

\author{Hanjun Wei}
\affiliation{%
  \institution{University of Chinese Academy of Sciences}
  \country{China}
}

\author{Zhe Cui}
\affiliation{%
  \institution{Chengdu Institute of Computer Applications, Chinese Academy of Sciences and University of Chinese Academy of Sciences}
  \country{China}
}

\author{Shiwen Ni}
\affiliation{%
  \institution{Artificial Intelligence Research Institute, Shenzhen University of Advanced Technology}
  \country{China}
}

\begin{abstract}
Large language models can generate functionally useful code that remains vulnerable, while security-oriented interventions can make code appear safe by breaking its intended behavior. We study a different intervention point: using security tests as executable specifications before generation and as feedback during repair. We develop \system, a controlled test-feedback scaffold that separates three decisions often conflated in prior work: whether tests are shown upfront, whether failed executions trigger revision, and how failures are selected and represented. The evaluation uses behavior-partitioned visible and hidden tests and byte-identical initial candidates for repair comparisons.

Across \PrimaryTrajectories{} trajectories, \PrimaryTaskInstances{} task instances, three secure-code benchmarks, \PrimaryCWEs{} CWE categories, and two model families, showing all visible tests upfront changes hidden functional-and-security joint success by +\UpfrontMacroDelta{} percentage points on average, but helps only \UpfrontPositive{} of nine benchmark--model conditions and hurts \UpfrontNegative{}. In shared-candidate comparisons, structured feedback repairs \MRepairs{} initially unsuccessful candidates with \MRegressions{} joint regressions; fixed raw feedback repairs \RawRepairs{} but regresses \RawRegressions{}. Yet structured and raw feedback are nearly indistinguishable head-to-head (\StructuredWins{} wins, \StructuredLosses{} losses, \StructuredTies{} ties). Candidates that pass all visible tests still fail hidden behavior families under every common regime. These findings support a mechanism-level conclusion rather than a winner claim: executable feedback can repair secure-code generation across model families, but its opportunity is bounded by test coverage, and its effect depends on the model, task, and feedback entry point. We release prompts, partitioned oracles, sandbox runners, complete trajectories, and frozen analysis manifests.
\end{abstract}

\keywords{secure code generation, large language models, test feedback, software testing, program repair, empirical study}

\maketitle

\section{Introduction}

Code generation is useful only when the generated program both implements its requirement and resists the relevant attacks. These goals are not interchangeable. A candidate that omits dangerous behavior can look secure while being useless; a candidate that passes ordinary unit tests can still contain command injection, path traversal, or improper certificate validation. Recent secure-code benchmarks therefore advocate evaluating functionality and security on the \emph{same} candidate~\cite{fu2024codeguard,peng2025cweval,dai2026rethinking}. Nevertheless, most defenses are still presented as model-training, constrained-decoding, static-analysis, or prompt-optimization techniques~\cite{he2023sven,he2024safecoder,nazzal2024promsec,nunez2024autosafecoder}. Less is known about a mechanism already familiar to developers: executing security tests during generation and returning failures to the model.

Tests can affect generation in at least three distinct ways. First, test code can be an \emph{upfront executable specification}, exposing examples and security properties before a candidate exists. Second, test execution can create a \emph{feedback loop} that revises an existing candidate. Third, an orchestrator can select and represent failures---for example, sending all raw logs or prioritizing a compact security failure. Prior studies of self-debugging and patch validation show that execution feedback can improve functional generation and vulnerability repair~\cite{chen2023selfdebug,le2022coderl,kulsum2024vrpilot}. However, a secure-generation claim requires stricter controls. If two methods generate different initial programs, an observed difference cannot be attributed to repair. If stopping uses hidden tests, the evaluation leaks its oracle. If visible and hidden tests exercise the same inputs, a reported gain may be test memorization. Finally, reporting security independently from functionality can reward empty or broken programs.

This paper presents a controlled empirical study of security tests at generation time. We build \system, a test-feedback scaffold, not as another universally superior prompting recipe, but as an instrument for separating the effects above. It runs candidate code in a sandbox, observes only a visible behavior partition, and performs at most two revisions. Fixed raw, random raw, and structured security-priority feedback reuse the exact initial response produced by the requirement-only condition. A second feedback condition similarly reuses the candidate generated with all visible tests upfront. Hidden functional and security tests run once, after stopping, and never enter a prompt or decision.

We evaluate \system on audited subsets of CWEval~\cite{peng2025cweval}, SALLM~\cite{siddiq2024sallm}, and CodeGuard+~\cite{fu2024codeguard}. The primary evidence contains \PrimaryTrajectories{} trajectories over 31 benchmark task instances and 16 CWE categories. It spans four local Qwen configurations from 7B to 32B and the remotely served DeepSeek-V4-Flash model, with five repetitions per condition. For external tasks whose original repositories did not provide the required partitioned dynamic oracle, we construct an overlay using a secure implementation, a functionality-preserving vulnerable mutant, repeat execution, behavior-family separation, and blinded machine semantic review. After the executable gates and three-repeat stability checks passed, three software-engineering doctoral students manually reviewed the frozen qualification packets. We therefore treat these external overlays as human-reviewed while retaining machine-assisted construction as a limitation.

The results complicate a simple ``more tests are better'' story. All tests upfront improve the benchmark--model macro average by 19.3 percentage points, but the direction is negative on Qwen2.5-Coder-7B-Instruct/SALLM and DeepSeek-V4-Flash/CodeGuard+. Shared-candidate feedback provides stronger causal evidence: structured feedback repairs 80 failures with no joint regression, while fixed raw feedback repairs 83 with three. Yet the structured strategy does not dominate raw feedback: across 465 paired cells, each wins six times. The breadth CodeGuard+ wave further shows why pilot effects can shrink. On 11 tasks selected without looking at baseline outcomes, feedback triggers in only 21 of 110 B0 cells and produces four repairs; a strong candidate that already passes visible tests offers no feedback opportunity even when it fails a hidden attack family.

This work makes four contributions:

\begin{itemize}[leftmargin=*]
  \item A controlled decomposition of upfront tests, executable feedback, and feedback representation for secure code generation, including shared-initial-candidate comparisons.
  \item A multi-benchmark, cross-family study comprising 2,705 trajectories, with functionality and dynamic security evaluated jointly on hidden behavior partitions.
  \item Mechanism-oriented measurements---repair/regression transitions and test-feedback overfitting---that explain when aggregate success rates rise or fall.
  \item A reproducibility package containing frozen prompts, model/configuration manifests, qualified oracles, sandbox executions, raw model responses, and analysis scripts.
\end{itemize}

Our conclusion is deliberately scoped: \emph{test feedback improves secure code generation by repairing a subset of failed candidates and makes the security--functionality trade-off observable; no tested feedback representation is best in all settings.}

\section{Background and Motivation}

\subsection{Joint secure-code evaluation}

Let $F(c)$ denote that candidate $c$ passes all functional tests and $S(c)$ that it passes all security tests. The useful outcome is $J(c)=F(c)\land S(c)$. Reporting only $S$ is unsafe as an evaluation practice: a program that returns a constant or deletes the requested operation may never reach a vulnerable sink. Reporting only $F$ misses vulnerability-specific behavior. CodeGuard+ introduced metrics that combine these dimensions~\cite{fu2024codeguard}; CWEval uses outcome-driven functional and security oracles~\cite{peng2025cweval}; and Dai et al. show that several apparent security improvements disappear when both properties are checked on the same output~\cite{dai2026rethinking}.

The strict conjunction is appropriate for deployability but hides partial functional progress. We therefore retain it as the primary outcome and report SAFE@1 as a sensitivity measure. SAFE weights security by an exponential transform of the functional test-case pass fraction~\cite{dai2026rethinking}. In our adaptation, ``secure'' means passing all task-specific dynamic security tests rather than receiving no warning from a collection of static analyzers. This difference matters and prevents a direct numerical comparison with the original SAFE study.

\subsection{Tests as specifications and feedback}

Test-driven development treats executable examples as a way to clarify intended behavior before implementation~\cite{beck2002tdd}. For an LLM, visible tests can likewise reduce ambiguity. Yet test source also consumes context, can anchor a model on particular values, and can shift attention from the prose contract. Upfront tests thus change the generation distribution; they do not demonstrate that a particular candidate was repaired.

Execution feedback is a different intervention. Self-debugging and CodeRL use test outcomes to guide additional generation~\cite{chen2023selfdebug,le2022coderl}; vulnerability-repair systems return compiler, functional, and sanitizer output~\cite{kulsum2024vrpilot}; secure-generation systems may loop over static analysis or fuzzing~\cite{nazzal2024promsec,nunez2024autosafecoder}. The result depends on what the oracle exposes. A feedback loop cannot react when every visible test passes, even if a hidden property fails. This creates a coverage-limited treatment opportunity that aggregate final scores alone do not show.

\subsection{The causal ambiguity of iterative generation}

Suppose method $A$ prompts a model once and method $B$ prompts it with tests, executes the output, and repairs it. If their initial outputs differ, the final contrast mixes prompt effects and repair effects. We avoid this ambiguity for the principal feedback comparisons: B4, B5, and M load the serialized B0 response rather than regenerating it; B6 similarly loads B3. Model, task, repetition, initial code, and initial usage accounting are therefore paired. Only the repair policy differs. B3 versus B0 remains useful, but we interpret it as a benchmark--model condition contrast, not an edit-level causal effect.

\section{SecTDD Experimental Scaffold}

\begin{figure}[t]
\centering
\resizebox{\linewidth}{!}{%
\begin{tikzpicture}[
  font=\small,
  node distance=5mm,
  stage/.style={
    draw=black!65,
    rounded corners=2pt,
    line width=0.55pt,
    minimum height=11mm,
    inner sep=3.5pt,
    align=center,
    fill=white
  },
  input/.style={stage,draw={rgb,255:red,0;green,119;blue,187},fill={rgb,255:red,232;green,244;blue,251}},
  visible/.style={stage,draw={rgb,255:red,0;green,153;blue,136},fill={rgb,255:red,229;green,247;blue,243}},
  feedback/.style={stage,draw={rgb,255:red,238;green,119;blue,51},fill={rgb,255:red,253;green,239;blue,229}},
  hidden/.style={stage,dashed,draw=black!70,fill=black!4},
  flow/.style={-{Latex[length=2.1mm,width=1.4mm]},semithick},
  note/.style={draw=black!45,rounded corners=2pt,fill=black!2,align=left,font=\scriptsize,inner sep=3.5pt}
]
  \node[input,text width=24mm] (task) {\textbf{Task}\\[-1pt]\scriptsize contract and visible oracle};
  \node[input,text width=29mm,right=of task] (prompt) {\textbf{Initial context}\\[-1pt]\scriptsize requirement $\pm$ visible tests};
  \node[stage,text width=21mm,right=of prompt] (candidate) {\textbf{LLM}\\[-1pt]\scriptsize candidate};
  \node[visible,text width=27mm,right=of candidate] (visible) {\textbf{Visible sandbox}\\[-1pt]\scriptsize $F^{\mathrm{vis}},S^{\mathrm{vis}}$};
  \node[stage,text width=23mm,right=of visible] (final) {\textbf{Freeze}\\[-1pt]\scriptsize final candidate};

  \draw[flow] (task) -- (prompt);
  \draw[flow] (prompt) -- (candidate);
  \draw[flow] (candidate) -- (visible);
  \draw[flow] (visible) -- node[above,font=\scriptsize] {pass / stop} (final);

  \node[feedback,text width=38mm,below=10mm of visible,xshift=-8mm] (feedback) {\textbf{Failure feedback}\\[-1pt]\scriptsize fixed raw / random raw / structured\\[-1pt]\scriptsize replace candidate\\[-1pt]\scriptsize at most two revisions};
  \draw[flow] (visible.south) -- node[right,font=\scriptsize] {fail + budget} (feedback.north);
  \draw[flow] (feedback.west) -| (candidate.south);

  \node[hidden,text width=28mm,below=10mm of final] (hidden) {\textbf{Hidden sandbox}\\[-1pt]\scriptsize final evaluation only};
  \node[stage,text width=28mm,below=5mm of hidden] (outcome) {\textbf{Hidden outcomes}\\[-1pt]\scriptsize $F^{\mathrm{hid}},S^{\mathrm{hid}},J$};
  \draw[flow] (final.south) -- node[right,font=\scriptsize] {run once} (hidden.north);
  \draw[flow] (hidden) -- (outcome);

  \node[note,text width=56mm,below=11mm of prompt] (pairing) {\textbf{Shared round-0 response.} B4/B5/M load the byte-identical B0 response; B6 loads B3 before the feedback branch.};
\end{tikzpicture}
}%
\Description{SecTDD generates a candidate from the task contract and optional visible tests, runs visible functional and security tests, and loops through bounded failure feedback for at most two revisions. The final candidate is frozen before hidden functional and security evaluation. A note identifies which methods share the same round-zero response.}
\caption{Controlled SecTDD workflow. Generation and revision observe only the visible partition. Hidden tests run once on the frozen candidate and never affect prompting, feedback, or stopping. The shared-response pairing isolates repair effects from differences in initial generation.}
\label{fig:workflow}
\end{figure}
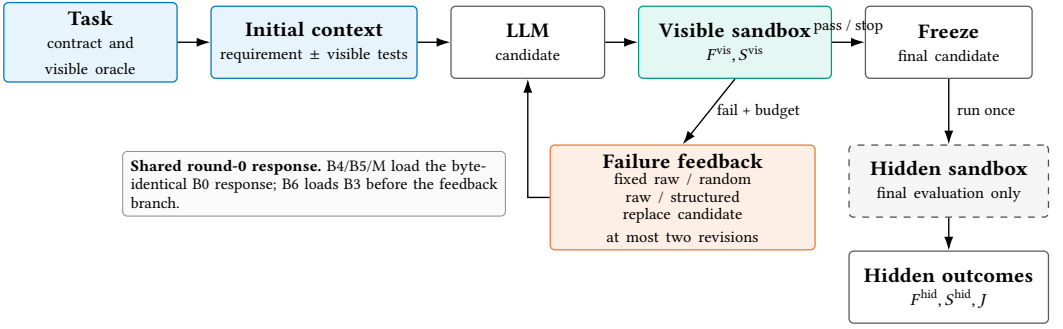

Figure~\ref{fig:workflow} shows the workflow. A task provides a natural-language contract and a behavior-partitioned oracle. The method controls what is placed in the first prompt. The candidate runs against visible functional and security cases in a fresh sandbox. If the method permits feedback and a visible case fails, the orchestrator constructs a bounded message and requests a complete replacement module. Stopping depends only on visible results and frozen resource limits. The hidden partition is mounted only for final evaluation.

\subsection{Feedback policies}

The structured policy first prioritizes failing security cases, then cases not previously selected, and finally a stable case identifier. At most two failures are returned per round. Each block contains the case identifier, marker (functional or security), an executable-property description, and the tail of the observed failure. Fixed raw feedback returns all failing logs in stable order. Random raw feedback returns one failure selected deterministically from the repetition identifier and round. All policies request one complete Python module; none reveal hidden inputs or results.

This design is intentionally modest. It tests whether selection and representation matter under a small, auditable budget. It does not claim that the fixed priority is an optimal scheduler. Indeed, the head-to-head results later show that representation effects are sparse relative to the effect of receiving any executable failure.

\subsection{Methods and controlled contrasts}

Table~\ref{tab:methods} defines the method family. Not every benchmark includes every diagnostic baseline. B1 and B2 appear only in CWEval; B5 appears in CWEval and the CodeGuard+ breadth wave; B6 appears on the external benchmarks. In the final operational protocol, B6 means ``B3 plus structured feedback.'' This supersedes an early living-protocol label that used B6 for a static-tool loop.

\begin{table}[t]
\caption{Experimental methods. M denotes the structured SecTDD policy. Iterative methods permit at most two revisions; ``Reused from'' identifies the condition supplying the byte-identical round-0 response.}
\label{tab:methods}
\centering
\begin{tabular}{llll}
\toprule
ID & Initial context & Feedback & Reused from \\
\midrule
B0 & Requirement & None & -- \\
B1 & Requirement + reminder & None & -- \\
B2 & Requirement + functional tests & None & -- \\
B3 & Requirement + all visible tests & None & -- \\
B4 & Requirement & All raw failures, fixed order & B0 \\
B5 & Requirement & One random raw failure & B0 \\
M  & Requirement & Structured security priority & B0 \\
B6 & Requirement + all visible tests & Structured security priority & B3 \\
\bottomrule
\end{tabular}
\end{table}

The principal controlled contrasts are B4$-$B0 and M$-$B0 (does feedback repair a requirement-only candidate?), B6$-$B3 (does feedback repair an upfront-test candidate?), and M$-$B4/B5 (does structured selection outperform raw feedback?). B3$-$B0 addresses the separate question of upfront specification.

\paragraph{Comparison scope.}
Training-time defenses such as SVEN and SafeCoder and decoder-specific defenses such as CodeGuard+ cannot be applied uniformly to the local and API checkpoints in our grid~\cite{he2023sven,he2024safecoder,fu2024codeguard}. Reimplementing them for only a subset would mix model and intervention effects. We therefore compare inference-time policies that operate through the same model interface and explicitly do not claim to outperform those systems. Their published findings motivate our joint evaluation; our question is orthogonal: under a fixed generator and executable oracle, what comes from test availability and feedback?

\section{Study Design}

We organize the study around five research questions:

\begin{description}[leftmargin=3.8em,style=nextline]
  \item[RQ1] How do visible security tests used as upfront specifications affect hidden joint success?
  \item[RQ2] Can executable feedback repair failed initial candidates without regressing functionality or security?
  \item[RQ3] Does structured security-priority feedback improve over raw feedback under shared initial candidates?
  \item[RQ4] How often does visible joint success fail to transfer to hidden behavior families?
  \item[RQ5] Which findings reproduce across benchmarks and model families?
\end{description}

\subsection{Benchmarks and task selection}

\paragraph{CWEval.}
CWEval provides multilingual security-critical generation tasks and outcome-driven functional/security evaluation~\cite{peng2025cweval}. We use nine Python tasks spanning eight CWE categories. These tasks participated in method development, so we label CWEval \emph{internal confirmatory} evidence rather than an untouched external test set.

\paragraph{SALLM.}
SALLM contains security-centric Python prompts derived from author examples, Stack Overflow, and CodeQL patterns~\cite{siddiq2024sallm}. Eleven tasks, each from a different CWE category, pass our executable-oracle qualification. SALLM is external to method development.

\paragraph{CodeGuard+.}
CodeGuard+ was designed to evaluate correctness and security jointly~\cite{fu2024codeguard}. Docker was unavailable in our execution environment, so we built a Python-only, Bubblewrap-based~\cite{bubblewrap} oracle overlay. The primary breadth wave is the union of tasks accepted by four frozen semantic panels: 11 tasks across eight CWE categories. Crucially, it does \emph{not} require a baseline security failure, a feedback trigger, or a pilot repair. This removes the outcome-selection filter present in an earlier three-task technical pilot, which we exclude from primary aggregates.

Across benchmarks there are 31 task instances and 16 distinct CWE identifiers from the MITRE taxonomy~\cite{mitrecwe}. Overlap in CWE number does not imply duplicate behavior: benchmark contracts and implementations remain separate.

\subsection{Oracle construction and qualification}

Each task has visible and hidden functional and security partitions. We split by behavior family or attack mechanism rather than randomly splitting individual assertions. A hidden case may preserve the security property while changing delimiters, encodings, nesting, boundary values, or attack construction. The hidden source, concrete inputs, expected outputs, and logs are unavailable to the generation process.

For SALLM and CodeGuard+ overlays, qualification has six gates:

\begin{enumerate}[leftmargin=*]
  \item A secure implementation must pass all functional and security cases.
  \item A vulnerability mutant must preserve functionality and fail at least one security case.
  \item Both subjects run three times in independent fresh sandboxes; every observation must be stable.
  \item Visible and hidden partitions must represent the declared behavior families without sharing concrete hidden material.
  \item Blinded semantic-review packets contain the contract, implementations, and partition descriptions but hide task provenance and construction labels. Two strong models review each packet; disagreements receive a third-model adjudication.
  \item After the executable and machine-review gates pass, three software-engineering doctoral students manually review each frozen packet for contract--oracle agreement, the semantics of the secure implementation and vulnerable mutant, attack realism, and visible/hidden partition separation.
\end{enumerate}

Strong models and the DeepSeek-V4-Flash API were also used to generate multiple candidate secure implementations during oracle construction. Candidates were accepted only by executable gates, not by model self-assertion. The use of generative AI in data/oracle construction is therefore part of the research method and is disclosed here. Model review served as a qualification aid rather than the sole semantic authority: after the executable gates and three-repeat stability checks, three software-engineering doctoral students manually reviewed every frozen SALLM and CodeGuard+ qualification packet. We therefore describe these overlays as human-reviewed, while carrying the limitations of machine-assisted construction through the paper.

\subsection{Models and generation}

Table~\ref{tab:scope} reports the exact model identifiers and complete primary grid. The local checkpoints are Qwen2.5-Coder-7B-Instruct and Qwen2.5-Coder-14B-Instruct~\cite{hui2024qwen25coder}, Qwen3-32B~\cite{yang2025qwen3}, and Qwen3.6-27B~\cite{qwen2026qwen36}. All run in BF16 with single-GPU inference. Thinking is disabled for Qwen3-32B and Qwen3.6-27B to keep code emission within the common budget. The cross-family conditions use the official API model DeepSeek-V4-Flash (provider identifier in Table~\ref{tab:scope}); thinking is explicitly disabled in every request. DeepSeek reports 284B total and 13B active parameters for V4-Flash, while the provider-managed server revision is not exposed~\cite{deepseek2026v4}. We therefore archive raw responses, request dates, usage, the provider identifier, and request slots instead of claiming bitwise reproducibility for the API condition.

\begin{table}[t]
\caption{Models and primary evidence grid. Panel A maps compact model keys to official identifiers and execution modes; Panel B enumerates the primary evaluation grid. Meth. denotes methods, Rep. denotes fixed seeds for local checkpoints and independent request slots for the remote API, and Traj. denotes trajectories. Repeated benchmark tasks and task--repetition cells are not independent inference units.}
\label{tab:scope}
\centering
\small
\begin{minipage}{0.92\textwidth}
{\raggedright\textbf{Panel A: Model registry}\par}
\vspace{0.25em}
\centering
\begin{tabular}{@{}lll@{}}
\toprule
Key & Official model or API identifier & Execution \\
\midrule
Q2.5-C7B  & \texttt{Qwen/Qwen2.5-Coder-7B-Instruct}  & Local, BF16 \\
Q2.5-C14B & \texttt{Qwen/Qwen2.5-Coder-14B-Instruct} & Local, BF16 \\
Q3-32B     & \texttt{Qwen/Qwen3-32B}                 & Local, BF16, no-thinking \\
Q3.6-27B   & \texttt{Qwen/Qwen3.6-27B}               & Local, BF16, no-thinking \\
DS-V4-F    & DeepSeek-V4-Flash (\texttt{deepseek-v4-flash}) & API, no-thinking \\
\bottomrule
\end{tabular}

\vspace{0.6em}
{\raggedright\textbf{Panel B: Primary evaluation grid}\par}
\vspace{0.25em}
\resizebox{\linewidth}{!}{\begin{tabular}{llrrrr}
\toprule
Benchmark & Model key & Tasks & Meth. & Rep. & Traj. \\
\midrule
CWEval & Q2.5-C7B & 9 & 7 & 5 & 315 \\
CWEval & Q2.5-C14B & 9 & 7 & 5 & 315 \\
CWEval & Q3-32B & 9 & 7 & 5 & 315 \\
SALLM & Q2.5-C7B & 11 & 5 & 5 & 275 \\
SALLM & Q2.5-C14B & 11 & 5 & 5 & 275 \\
SALLM & Q3.6-27B & 11 & 5 & 5 & 275 \\
SALLM & DS-V4-F & 11 & 5 & 5 & 275 \\
CodeGuard+ & Q2.5-C14B & 11 & 6 & 5 & 330 \\
CodeGuard+ & DS-V4-F & 11 & 6 & 5 & 330 \\
\midrule
\multicolumn{5}{r}{Total} & 2,705 \\
\bottomrule
\end{tabular}
}
\end{minipage}
\end{table}

Temperature is 0.2. CWEval permits 2,048 completion tokens per call; external experiments permit 3,072. Iterative methods make at most three calls, use at most 1,200 feedback tokens total (600 per round), and stop when all visible tests pass or a call, token, execution, or timeout budget is reached. CWEval has a 12,000-input/6,144-output-token total budget and at most 20 visible executions; the external grids use 14,000/9,216 tokens and at most six visible executions. These limits were frozen before each confirmatory wave.

Local repetitions use five fixed generation seeds. The DeepSeek-V4-Flash endpoint did not support deterministic seeds: its five identifiers denote independent request slots and pairing keys and were not sent as provider seeds. B4/B5/M still reuse the exact B0 response, and B6 the exact B3 response, within each slot.

\subsection{Outcomes}

For task $x$, repetition $r$, and final candidate $c_{xr}$, the primary metric is
\begin{equation}
\textsc{HSP@1}=\frac{1}{NR}\sum_{x=1}^{N}\sum_{r=1}^{R}
\mathbb{1}\!\left[F^{hid}(c_{xr})\land S^{hid}(c_{xr})\right].
\end{equation}
We call it \joint in prose. We additionally report \func, \secure, security-given-functional $P(S^{hid}=1\mid F^{hid}=1)$, SAFE@1, and the four states $F^+S^+$, $F^+S^-$, $F^-S^+$, and $F^-S^-$. Security alone is never used as the headline result.

For feedback, we count a \emph{trigger} when an iterative condition makes at least one revision, a \emph{joint repair} when its shared initial candidate fails hidden joint evaluation and its final candidate passes, and a \emph{joint regression} for the reverse transition. We separately count functionality regressions and function-preserving security repairs ($F^+S^-\rightarrow F^+S^+$).

To quantify coverage-limited stopping, we define the test-feedback overfitting rate:
\begin{equation}
\textsc{TFOR}=\frac{\sum_{x,r}\mathbb{1}[J^{vis}(c_{xr})\land\neg J^{hid}(c_{xr})]}
{\sum_{x,r}\mathbb{1}[J^{vis}(c_{xr})]}.
\end{equation}
\tfor is an operational diagnostic for this study, not a claim that no equivalent generalization metric exists. We report its numerator and denominator because a small denominator makes the rate unstable.

\subsection{Statistical analysis and integrity checks}

The task is the inference unit; the 2,705 trajectories are repeated observations, not independent samples. We compute method rates over task--repetition cells and 95\% confidence intervals using 10,000 task-cluster bootstrap resamples~\cite{efron1993bootstrap}. Confirmatory method differences use exact sign-flip permutation tests over task-level mean differences and Holm correction within each frozen comparison family~\cite{holm1979}. Given the small task sets, we emphasize effect sizes and paired events. The smallest adjusted $p$ in component reports is \MinimumHolmP; no comparison reaches $0.05$.

Every run manifest records task, model, method, repetition, budgets, prompt and oracle hashes, candidate hashes, token usage, and execution results. Analysis aborts on a missing grid cell, model mismatch, hash drift, or unexpected trajectory count. A hidden-leak audit scans prompts and feedback for hidden identifiers, descriptions, and non-public string literals. Code runs without network access in a fresh minimal mount with per-test and wall-clock timeouts.

\section{Results}

Table~\ref{tab:joint} gives the principal outcome for every condition. Percentages are deliberately not pooled across heterogeneous benchmark grids.

\begin{table}[t]
\caption{\joint (\%) by benchmark, model key, and method. Model keys are defined in Table~\ref{tab:scope}; dashes denote methods not run on that benchmark.}
\label{tab:joint}
\centering
\resizebox{\textwidth}{!}{\begin{tabular}{llrrrrrrrr}
\toprule
Benchmark & Model key & B0 & B1 & B2 & B3 & B4 & B5 & M & B6 \\
\midrule
CWEval & Q2.5-C7B & 15.6 & 22.2 & 33.3 & 40.0 & 48.9 & 46.7 & 48.9 & -- \\
CWEval & Q2.5-C14B & 28.9 & 44.4 & 37.8 & 57.8 & 57.8 & 48.9 & 53.3 & -- \\
CWEval & Q3-32B & 22.2 & 11.1 & 33.3 & 73.3 & 62.2 & 57.8 & 62.2 & -- \\
SALLM & Q2.5-C7B & 45.5 & -- & -- & 41.8 & 60.0 & -- & 63.6 & 58.2 \\
SALLM & Q2.5-C14B & 54.5 & -- & -- & 63.6 & 70.9 & -- & 70.9 & 72.7 \\
SALLM & Q3.6-27B & 60.0 & -- & -- & 90.9 & 69.1 & -- & 69.1 & 90.9 \\
SALLM & DS-V4-F & 52.7 & -- & -- & 81.8 & 67.3 & -- & 67.3 & 87.3 \\
CodeGuard+ & Q2.5-C14B & 76.4 & -- & -- & 87.3 & 78.2 & 78.2 & 78.2 & 89.1 \\
CodeGuard+ & DS-V4-F & 89.1 & -- & -- & 81.8 & 94.5 & 94.5 & 94.5 & 81.8 \\
\bottomrule
\end{tabular}
}
\end{table}

\subsection{RQ1: Upfront tests are helpful, but not monotonic}

B3 exceeds B0 in seven of nine benchmark--model conditions and is worse in two. The unweighted benchmark--model macro difference is +\UpfrontMacroDelta{} percentage points. The largest improvements occur on CWEval/Qwen3-32B (22.2\% to 73.3\%), SALLM/Qwen3.6-27B (60.0\% to 90.9\%), and SALLM/DeepSeek-V4-Flash (52.7\% to 81.8\%). These results support the idea that visible tests can convey security-relevant semantics more concretely than prose alone.

The exceptions reveal the risk. On SALLM/Qwen2.5-Coder-7B-Instruct, B3 falls from 45.5\% to 41.8\%, driven by a secure-rate reduction from 74.5\% to 56.4\% even as functionality rises. On CodeGuard+/DeepSeek-V4-Flash, B3 falls from 89.1\% to 81.8\%. In the latter condition, the all-tests prompt produces five $F^+S^-$ cells and a \tfor of 18.2\%, versus two $F^+S^-$ cells and 0\% TFOR for B0. A post-hoc CWE-89 example is especially instructive: the B3 candidate passes the complete visible partition but fails the hidden data-integrity behavior, while the same-slot B0 candidate passes hidden joint evaluation. Because the candidates are independently generated, this example illustrates a prompt-distribution effect rather than a code edit.

B1 and B2 on CWEval further argue against a universal prompt heuristic. A generic security reminder improves Qwen2.5-Coder-14B-Instruct from 28.9\% to 44.4\% but harms Qwen3-32B from 22.2\% to 11.1\%. Functional tests alone improve all three CWEval models over B0, yet remain below all-tests B3. Thus, executable security examples often add useful information, but model capacity and attention allocation determine whether that information is used safely.

\noindent\fbox{\parbox{0.96\textwidth}{\textbf{Answer to RQ1.} Upfront visible tests usually improve hidden joint success (7/9 conditions; +19.3 points macro), but can reduce it. They are a model-dependent specification mechanism, not a monotonic safety intervention.}}

\subsection{RQ2: Feedback repairs shared candidates}

Table~\ref{tab:feedback} aggregates only comparisons with shared initial responses. M triggers on 252 of 465 B0 cells and turns 80 hidden joint failures into successes, with no joint regression. Forty of those are function-preserving security repairs. Fixed raw feedback produces 83 repairs and three joint regressions. B6 triggers less often---47 of 330 B3 cells---but repairs 18 and has no joint regression. It does incur one security regression that does not cross from joint success to joint failure because the B3 reference was not jointly successful.

\begin{table}[t]
\caption{Paired feedback transitions over shared initial candidates. Task--rep. cells are repeated observations, not independent inference units. Repair denotes hidden joint failure $\rightarrow$ success; Trig. denotes at least one revision; J.Reg./F.Reg. denote hidden joint/function regression; Sec.-only denotes $F^+S^-\rightarrow F^+S^+$.}
\label{tab:feedback}
\centering
\begin{tabular}{lrrrrrr}
\toprule
Comparison & Task--rep. & Trig. & Repair & J.Reg. & F.Reg. & Sec.-only \\
\midrule
B4 $-$ B0 & 465 & 252 & 83 & 3 & 15 & 45 \\
M $-$ B0 & 465 & 252 & 80 & 0 & 8 & 40 \\
B6 $-$ B3 & 330 & 47 & 18 & 0 & 0 & 2 \\
M $-$ B4 & 465 & 252 & 6 & 6 & 8 & 0 \\
\bottomrule
\end{tabular}

\end{table}

The opportunity differs sharply by benchmark. On internal CWEval, M triggers 113/135 cells and repairs 44, including 38 security-only failures, but also causes eight functionality regressions relative to B0. Because none of the regressed B0 candidates were hidden joint successes, the aggregate joint-regression count remains zero; the four-state decomposition is what exposes the trade-off. On SALLM, M triggers 118/220 and repairs 32 without a functionality regression. On outcome-unfiltered CodeGuard+, it triggers only 21/110 and repairs four. This shrinkage from internal to breadth evidence is not a contradiction: many CodeGuard+ candidates already pass the visible partition, so the feedback treatment is never delivered.

A representative CWEval CWE-79 trajectory begins at $F^+S^-$: the code implements the required rendering but fails hidden HTML entity and attribute-delimiter behaviors. One structured visible failure causes a second candidate that retains functionality and passes all hidden security cases. A cross-family SALLM CWE-327 trajectory begins at $F^-S^-$ and fails rejection of MD5 and a mismatching SHA-256 digest; one B6 revision reaches $F^+S^+$. Hidden behavior labels are used only in this post-hoc explanation, not in feedback.

Costs track treatment delivery. Relative to their shared reference, B4 uses 0.81 additional calls and about 178 feedback tokens per eligible cell on average; M uses 0.80 calls and 202 tokens; B6 uses 0.24 calls and 49 tokens because B3 more often passes visibly. Feedback therefore buys repairs at non-trivial but bounded inference cost.

\noindent\fbox{\parbox{0.96\textwidth}{\textbf{Answer to RQ2.} Executable feedback can causally repair a subset of failed shared candidates across all three benchmarks and both model families. It also exposes lower-level functionality/security regressions that a single joint score can hide.}}

\subsection{RQ3: Structure does not universally beat raw logs}

M and B4 share 465 initial candidates and differ only in failure selection/representation. M wins six cells, B4 wins six, and 453 tie. The wins and losses concentrate in CWEval and SALLM; all 110 CodeGuard+ cells tie. Aggregate repair counts slightly favor raw feedback (83 versus 80), while joint regressions favor structured feedback (zero versus three). These differences are too sparse to support a universal ranking.

The qualitative failures explain why. In one SALLM CWE-22 slot, fixed raw feedback reaches $F^+S^+$ after one revision. Structured feedback continues to a third round and ends at $F^-S^+$, failing hidden nested-file deletion. Prioritizing security reduced the immediate vulnerability signal but did not preserve the complete functional contract. Conversely, structured selection can avoid overwhelming the prompt when many raw failures are redundant. Which behavior dominates depends on the failure set and model.

In the CodeGuard+ breadth wave, B4, B5, and M have identical joint outcomes for both models: 78.2\% on Qwen2.5-Coder-14B-Instruct and 94.5\% on DeepSeek-V4-Flash. They also make the same repair transitions. This equality is substantive: a complex representation policy has no room to differentiate when few failures trigger and the selected messages lead to the same code-level correction.

\noindent\fbox{\parbox{0.96\textwidth}{\textbf{Answer to RQ3.} We find no outcome-level superiority for structured feedback over raw feedback. The robust effect belongs to receiving executable feedback; representation effects are conditional and sparse.}}

\subsection{RQ4: Visible success leaves a coverage gap}

Table~\ref{tab:tfor} computes \tfor for the four methods available on every primary benchmark grid. Between 5.6\% and 18.8\% of visible-joint candidates fail hidden joint evaluation. B3 has the largest pooled rate despite strong final success in several conditions. This does not mean B3 is always worse: TFOR is conditional on visible success and should be read beside overall \joint.

\begin{table}[t]
\caption{Visible-to-hidden generalization for methods available on all benchmark grids. Hidden-joint failures are counted among visible-joint passes. Counts pool heterogeneous tasks for diagnosis, not significance testing.}
\label{tab:tfor}
\centering
\begin{tabular}{lrrr}
\toprule
Method & Visible-joint passes ($n$) & Hidden-joint failures ($n$) & TFOR (\%) \\
\midrule
B0 & 213 & 12 & 5.6 \\
B3 & 389 & 73 & 18.8 \\
B4 & 360 & 49 & 13.6 \\
M & 360 & 50 & 13.9 \\
\bottomrule
\end{tabular}

\end{table}

The failure occurs under every common regime. For example, a CWEval URL-validation candidate passes all visible checks but fails hidden deep-subdomain, authority-userinfo, and encoded-delimiter families. More importantly for iteration, DeepSeek-V4-Flash/CodeGuard+ B6 makes no revision in any of 55 cells: its B3 candidates all pass the visible partition, yet five fail hidden joint evaluation. This is not a failure of message formatting; no message exists. Increasing revision budget would not help unless the visible oracle exposes a distinguishing behavior.

Feedback can also increase visible confidence without proportionally improving hidden success. On Qwen2.5-Coder-14B-Instruct/CodeGuard+, B4/M move from B0's 9.8\% TFOR to 15.6\% while producing a small +1.8-point joint improvement over their shared candidates. Some repaired visible failures remain outside the hidden acceptance region. The correct engineering response is stronger and more diverse tests, not merely more rounds.

\noindent\fbox{\parbox{0.96\textwidth}{\textbf{Answer to RQ4.} Passing all visible functional and security tests is not a reliable safety certificate. Feedback is coverage-limited: it cannot repair a hidden property that produces no visible failure.}}

\subsection{RQ5: Effects reproduce qualitatively, not uniformly}

Table~\ref{tab:codeguard} focuses on the untouched, outcome-unfiltered CodeGuard+ breadth wave. Qwen2.5-Coder-14B-Instruct benefits from B3 over B0 (+10.9 points) and B6 over B3 (+1.8), while B4/M improve their shared B0 candidates by only +1.8. DeepSeek-V4-Flash shows the opposite entry-point pattern: B3 is 7.3 points below B0, B6 never triggers, and B4/M repair three B0 cells to gain 5.5 points. The same test-feedback mechanism exists in both families, but the useful starting point differs.

\begin{table}[t]
\caption{Outcome-unfiltered CodeGuard+ breadth results. Func., Secure, Joint, and TFOR are hidden-evaluation percentages; Calls is the mean number of model calls per task--repetition cell. Model keys are defined in Table~\ref{tab:scope}. This wave contains 11 human-reviewed overlay tasks and applies no baseline-failure or feedback-trigger filter.}
\label{tab:codeguard}
\centering
\resizebox{0.92\textwidth}{!}{\begin{tabular}{llrrrrr}
\toprule
Model key & Method & Func. & Secure & Joint & TFOR & Calls \\
\midrule
Q2.5-C14B & B0 & 80.0 & 80.0 & 76.4 & 9.8 & 1.00 \\
Q2.5-C14B & B3 & 89.1 & 90.9 & 87.3 & 8.5 & 1.00 \\
Q2.5-C14B & B4 & 87.3 & 89.1 & 78.2 & 15.6 & 1.44 \\
Q2.5-C14B & B5 & 87.3 & 83.6 & 78.2 & 15.6 & 1.44 \\
Q2.5-C14B & M & 87.3 & 90.9 & 78.2 & 15.6 & 1.44 \\
Q2.5-C14B & B6 & 90.9 & 92.7 & 89.1 & 8.3 & 1.27 \\
\midrule
DS-V4-F & B0 & 92.7 & 89.1 & 89.1 & 0.0 & 1.00 \\
DS-V4-F & B3 & 90.9 & 90.9 & 81.8 & 18.2 & 1.00 \\
DS-V4-F & B4 & 94.5 & 100.0 & 94.5 & 5.5 & 1.13 \\
DS-V4-F & B5 & 94.5 & 100.0 & 94.5 & 5.5 & 1.13 \\
DS-V4-F & M & 94.5 & 100.0 & 94.5 & 5.5 & 1.13 \\
DS-V4-F & B6 & 90.9 & 90.9 & 81.8 & 18.2 & 1.00 \\
\bottomrule
\end{tabular}
}
\end{table}

SALLM provides another cross-family check. For DeepSeek-V4-Flash, B6 improves B3 from 81.8\% to 87.3\%; for Qwen2.5-Coder-14B-Instruct it improves 63.6\% to 72.7\%. For Qwen3.6-27B, however, both are already 90.9\%, and for Qwen2.5-Coder-7B-Instruct B6 ends below M. Across Qwen and DeepSeek, feedback can repair failures; across model/benchmark cells, neither its magnitude nor the best entry condition is stable.

The outcome-unfiltered breadth result also changes the interpretation of the earlier selected pilot. A three-task CodeGuard+ pilot, selected after a preflight requiring a failure or feedback opportunity, showed large feedback gains. When all 11 semantically qualified tasks are included without looking at outcomes, triggers and gains shrink. We exclude the selected pilot from the 2,705 primary trajectories and use it only as evidence that preflight selection can inflate apparent treatment opportunity.

\noindent\fbox{\parbox{0.96\textwidth}{\textbf{Answer to RQ5.} The qualitative repair mechanism reproduces across Qwen and DeepSeek and across three benchmarks. Effect size, direction of upfront prompting, and the best feedback entry point do not.}}

\section{Discussion}

\subsection{What the study establishes}

The strongest evidence is paired and mechanistic. When a visible failure exists, executing tests and returning the failure can change the \emph{same} initial candidate from hidden failure to hidden joint success. We observe 80 such changes for structured B0 feedback and 18 for structured B3 feedback. The result appears on internal and external benchmarks and both model families.

The study does not establish that M is a new state-of-the-art method. Raw and structured feedback are essentially tied head-to-head, and no adjusted test is significant. Treating the study as an algorithm leaderboard would discard its main contribution: separating specification, feedback availability, representation, and coverage reveals why secure-generation interventions sometimes help and sometimes harm.

\subsection{Implications for tool builders}

\paragraph{Preserve the joint contract.}
An orchestrator should track both functional and security properties after every edit. Security-first selection can produce $F^-S^+$, while functionality-first correction can reintroduce a vulnerability. Reporting only the dimension targeted by the latest feedback creates false progress.

\paragraph{Measure treatment opportunity.}
A high final score may arise from a strong initial model rather than a feedback policy; a low feedback gain may arise because tests never fail. Tools should report trigger count, repairs per trigger, and the initial four-state distribution alongside final pass rates.

\paragraph{Invest in behavior diversity before more rounds.}
When visible tests pass and hidden attacks fail, extra iterations are inert. Mutation testing, metamorphic variants, boundary generation, and independent fuzzing can expand the visible signal. Care is needed to prevent the new tests from becoming near-duplicates of the hidden evaluation.

\paragraph{Use structured feedback for control, not assumed accuracy.}
Compact typed feedback has engineering advantages: predictable token use, explicit property labels, and prioritization. Our evidence does not show that it is inherently more effective than raw logs. A practical system could choose adaptively: structured summaries for redundant failure sets, raw context for subtle functional traces, and regression tests retained across rounds.

\subsection{Implications for secure-code evaluation}

First, function and security must be evaluated on the same candidate. Second, hidden tests must be isolated from generation and stopping. Third, iterative methods need shared-candidate analysis; otherwise initial prompt differences masquerade as repair. Fourth, task-level inference matters. Thousands of trajectories cannot compensate for a small number of security tasks. Finally, benchmark qualification should be separated from outcome selection. Our CodeGuard+ breadth wave demonstrates how a semantically valid but outcome-unfiltered set can yield a smaller, more credible effect.

\subsection{A research agenda}

The next step is not another fixed priority rule. A stronger line of work would learn or search feedback policies while evaluating on held-out tasks and behavior families; estimate the value of a test before spending a model call; generate new tests when the visible suite reaches a ceiling; and extend joint evaluation to repository-level changes, multiple languages, build systems, and stateful vulnerabilities. Human studies are also needed: SAFE@1 assumes partially functional secure code has developer value, but the actual effort to complete such code remains unmeasured in our study.

\section{Threats to Validity}

\paragraph{Construct validity.}
Passing dynamic tests does not prove absence of vulnerabilities. We cover specified CWE behaviors, not arbitrary attacks. Conversely, a security test may reject a safe but semantically different implementation. Secure implementations, vulnerable mutants, repeat executions, and semantic review reduce this risk but do not eliminate it. SAFE@1 often collapses to the joint outcome on external tasks because they contain one hidden functional case; we therefore treat SAFE as sensitivity analysis, not independent corroboration. TFOR depends on the strength and granularity of both partitions.

\paragraph{Internal validity.}
Shared initial candidates make B4/B5/M versus B0 and B6 versus B3 edit-level comparisons. B3 versus B0 is not shared and cannot identify a repair effect. Local generation seeds improve repeatability but do not remove stochastic inference variation. Remote request slots are not deterministic seeds, and the provider may revise its backend. Token accounting for DeepSeek-V4-Flash uses a frozen local proxy tokenizer for budget enforcement; raw provider usage is retained separately. Different benchmarks have slightly different completion and execution budgets, so we do not pool them as a single leaderboard.

\paragraph{Oracle and researcher bias.}
Generative models proposed secure implementations and served as preliminary semantic reviewers. Executable mutation gates prevent accepting candidates on model opinion alone, and review packets are blinded, but correlated model errors remain possible. After the executable gates and three-repeat stability checks, three software-engineering doctoral students manually reviewed every frozen SALLM and CodeGuard+ qualification packet. This human review reduces, but does not eliminate, construct and researcher bias: reviewers may share interpretation assumptions, and passing a finite dynamic oracle cannot establish security against arbitrary attacks.

\paragraph{External validity.}
The study covers two model families, four local size/configuration points, a remote API, three benchmarks, and 16 CWE categories. All tasks are Python functions or small components. Findings may not transfer to Java/C/C++, repository-level agents, multi-file state, concurrency, dependency vulnerabilities, or production deployment. Public benchmark prompts may occur in pretraining. Behavior-family splits reduce direct input memorization but cannot establish contamination-free evaluation.

\paragraph{Statistical conclusion validity.}
There are 31 benchmark task instances, not 2,705 independent samples. Cluster intervals are wide, and no Holm-adjusted comparison is below 0.05. The nine-condition RQ1 macro treats benchmark--model cells equally and is descriptive. Repair counts establish observed transitions, not a population-wide effect size. We avoid universal superiority language and expose denominators throughout.

\paragraph{Reliability and safety.}
Generated code is untrusted. Bubblewrap restricts filesystem mounts, process environment, and network access, with fresh sandboxes and timeouts, but an unprivileged sandbox is not equivalent to a hardened VM. The artifact should run only on an isolated research host. Local checkpoint hashes and all prompts/oracles are frozen; remote API outputs are archived because they may not be reproducible.

\section{Related Work}

\subsection{Secure code generation}

Early empirical work showed that code assistants can emit CWE-relevant vulnerabilities and that access to an AI assistant can worsen security outcomes while increasing users' confidence in their solutions~\cite{pearce2022asleep,perry2023insecure}. SVEN steers pretrained code models with property-specific continuous vectors~\cite{he2023sven}; SafeCoder jointly tunes instruction following and secure generation~\cite{he2024safecoder}; CodeGuard+ studies constrained decoding and introduces joint correctness/security evaluation~\cite{fu2024codeguard}; and PromSec iteratively optimizes prompts using vulnerability-analysis feedback~\cite{nazzal2024promsec}. AutoSafeCoder combines coding, static-analysis, and fuzzing agents~\cite{nunez2024autosafecoder}. These approaches differ in whether intervention occurs during training, decoding, prompting, or tool use. Our work intervenes only at inference and holds initial candidates fixed to isolate executable-feedback effects.

Dai et al. reevaluate secure-generation methods with multiple analyzers and functionality on the same code, showing that apparent safety can come from broken output~\cite{dai2026rethinking}. Our dynamic joint outcome follows the same principle and adds longitudinal state transitions. Their SAFE@k motivates our partial-functionality sensitivity analysis, while hidden dynamic security cases remain our primary oracle.

\subsection{Secure-code benchmarks}

SecurityEval established a CWE-mapped prompt set for security assessment of generated code~\cite{siddiq2022securityeval}; SALLM supplies security-centric Python prompts and configurable assessment~\cite{siddiq2024sallm}. SecCodePLT argues for executable security evaluation in realistic environments~\cite{yang2024seccodeplt}. CWEval provides outcome-driven functional and security oracles~\cite{peng2025cweval}, and CodeGuard+ explicitly measures secure and correct code~\cite{fu2024codeguard}. On functional generation, EvalPlus demonstrates that weak test suites can accept incorrect programs and even mis-rank models~\cite{liu2023evalplus}. We do not propose a replacement benchmark. We adapt three complementary sources into a common visible/hidden protocol and document the semantic qualification needed when original execution environments are unavailable.

\subsection{Execution feedback and repair}

CodeT generates tests to select among candidate programs, while LEVER learns to rank candidates from their execution results~\cite{chen2023codet,ni2023lever}. CodeRL incorporates unit-test feedback in training and inference~\cite{le2022coderl}; self-debugging revises generated programs using execution results~\cite{chen2023selfdebug}; and Reflexion and OpenCodeInterpreter broaden iterative linguistic or execution feedback to agentic code refinement~\cite{shinn2023reflexion,zheng2024opencodeinterpreter}. VRpilot returns compiler, functional, and sanitizer feedback for vulnerability repair~\cite{kulsum2024vrpilot}. Our focus differs in three respects: generated secure code rather than a known vulnerable program, a hidden joint functional/security endpoint, and exact sharing of initial candidates across feedback policies. The visible-to-hidden design also exposes the point at which iterative debugging cannot proceed because its oracle is silent.

\section{Conclusion}

Security tests can serve as more than a final benchmark. Used upfront, they often improve generation but sometimes harm it. Used as feedback, they repair a measurable subset of the same failed candidates across benchmarks and model families. The benefit is neither free nor universal: feedback can shift functionality and security in different directions, structured summaries do not consistently beat raw logs, and no loop can repair a hidden behavior that visible tests fail to expose. Secure-code generation should therefore be evaluated as a joint, budgeted, coverage-dependent process. The empirical claim supported by our data is not that one method wins everywhere; it is that executable feedback makes security repair possible and its limits measurable.

\section*{Data Availability}

An anonymized replication package is available to reviewers at \url{https://anonymous.4open.science/r/sectdd-FBDB/}. It contains the living and frozen protocols, benchmark manifests and repository commits, visible/hidden oracle projections, secure and vulnerable oracle subjects, sandbox runners, exact prompts, local checkpoint/configuration hashes, raw responses, the 2,705 trajectory records analyzed in this manuscript, analysis scripts, generated tables, and result freezes. Hidden tests are included in the reviewer package but are never mounted during generation. API credentials, absolute user paths, and identity-bearing metadata are excluded. Upon acceptance, the authors intend to publish the curated package under the licenses of the component benchmarks.

\bibliographystyle{ACM-Reference-Format}
\bibliography{references}

\end{document}